\documentclass[epj]{svjour}
\usepackage{graphics}
\begin{document}
\title{Analysis of CLAS data on double charged-pion
electroproduction.}
\author{V.~I.~Mokeev\inst{1,2} \and V.~D.~Burkert\inst{1} \and L.~Elouadrhiri\inst{1} \and
 G.~V.~Fedotov\inst{2}  \and  E.~N.~Golovach\inst{2} \and 
 E.~L.Isupov\inst{2}, B.~S.~Ishkhanov\inst{2} 
 \and N.~V.~Shvedunov\inst{2} \and CLAS Collaboration%
}                     
%
%
\institute{Jefferson Lab, 12000, Jefferson Ave, Newport News, VA, 23606, USA \and 
Skobeltsyn Nuclear Physics Institute at Moscow State University,119899, Leninskie gory,
Moscow, Russia.}
\date{Received: date / Revised version: date}
%
\abstract{Recent developments in phenomenological analysis of the CLAS data on 2$\pi$
electroproduction are presented. The contributions from isobar channels and 
$P_{11}(1440)$, $D_{13}(1520)$ electrocouplings at $Q^{2}$ from 0.25 to 0.6
GeV$^2$  were 
determined from the analysis of 
comprehensive data on differential and fully integrated 2$\pi$ cross sections.} 

\maketitle
\section{Introduction}
\label{intro}
Studies of nucleon resonance electrocouplings at various photon virtualities in double 
charged-pion electroproduction play an important role in the $N^{*}$ program 
with the CLAS detector
\cite{Bu06,Bu07}. Single and double pion photo and electroproduction are two major exclusive channels,
contributing to the total photon-proton cross section in the $N^{*}$ excitation 
region. Both of these
channels are sensitive to excited states. Photo and electroproduction 
of two pions are
particularly sensitive to resonances with masses above 1.6 GeV. Many of these
states decay
preferentially to final states with two pions. Furthermore, 1$\pi$ and 2$\pi$ exclusive
channels are strongly coupled by hadronic interactions
in the final states (FSI). Hadronic cross section $\pi N \rightarrow \pi \pi N$ 
is the second strongest exclusive channel in
value amongst exclusive $\pi N$ cross sections. Therefore, a combined analysis
of at least the two major electroproduction channels is needed to assure the appropriate evaluation of $N^{*}$
electrocouplings. Eventually other
exclusive channels with smaller cross sections may be included. 
For these final states, 
the hadronic
interactions with major meson photo or electroproduction channels become
even more important. Therefore, comprehensive
information on mechanisms contributing to both 1$\pi$ and 2$\pi$ 
electroproduction is of
particular interest for the entire $N^{*}$ program. This information may be 
obtained in 
a phenomenological 
analysis of the CLAS data on meson electroproduction offering valuable input for $N^{*}$
studies in advanced coupled channel approaches, which are currently under 
development at the  
Excited Baryon Analysis Center (EBAC) at JLAB \cite{Lee06}. 

In this proceeding we report results of an analysis of recent CLAS data on
double charged-pion electroproduction \cite{Fe07,Mo07} at $W$$<$ 1.6 GeV and $Q^{2}$ 
from 0.2 to
0.6 GeV$^2$. The objectives of this analysis were: 
\begin{itemize}
\item establish all significant contributing mechanisms;
\item determine the electrocouplings of $P_{11}(1440)$ and $D_{13}(1520)$ states.
\end{itemize}

\section{JM06 model for phenomenological data analysis.}
\label{sec:1}
\begin{figure}[htbp]
\vspace{17.0cm}
\includegraphics{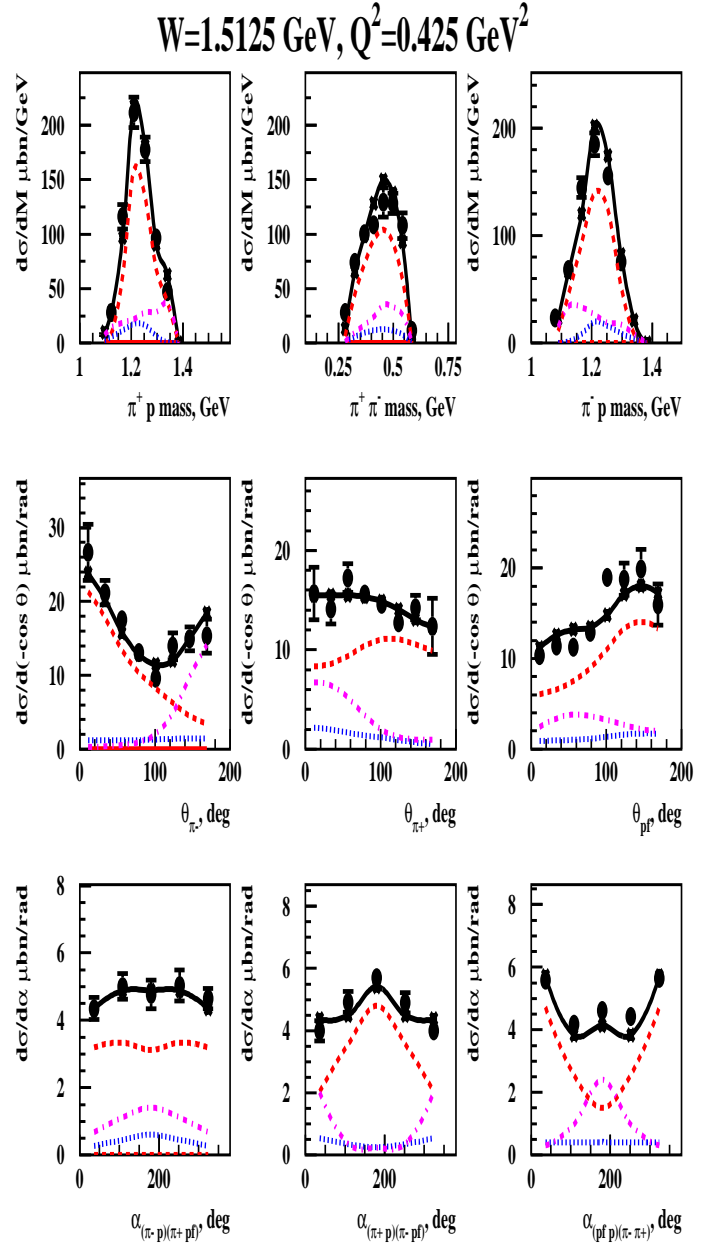} 
\caption[]{\small{Full set of differential 2$\pi$ cross sections obtained with CLAS 
at $W$=1.51 GeV and $Q^{2}$=0.425 GeV$^2$} in comparison with JM06 results
(solid lines). The contributions from $\pi^{-} \Delta^{++}$, $\pi^{+}
\Delta^{0}$ isobar channels and direct 2$\pi$ production mechanisms are shown 
by dashed, dotted  and dot-dashed lines, respectively.  }
\label{1diff}
\end{figure}

The exclusive $\gamma p \rightarrow \pi^{-} \pi^{+} p$ channel offers numerous 
observables for the 
analysis. Even in measurements 
without use of polarization observables, there are nine
independent differential cross sections in each ($W$,$Q^{2}$) bin. For the 
first time 
these data have become available from CLAS and some of them are shown in 
Fig.~\ref{1diff} and \ref{fitsec}. The full data set
obtained in the CLAS experiment \cite{Fe07,Mo07} may be found in Ref. \cite{db}.
These data make it possible to establish all significant contributing mechanisms from the
studies of their
manifestations in observables, as peaks in invariant mass distributions or 
sharp slopes in
angular distributions. Mechanisms without particular features may
be determined, considering the correlations between shapes of their cross sections in 
various observables. 
We have developed a phenomenological model that incorporates particular 
meson-baryon mechanisms
based on their manifestations in the observables. 

The analysis of earlier CLAS data \cite{Ri03} incorporated
 particular
meson-baryon mechanisms needed to describe $\pi^{+}$p, $\pi^{+}$$\pi^{-}$,
$\pi^{-}$p invariant masses and
$\pi^{-}$ angular distribution \cite{Mo01,Mo03,Mo07-1}. These cross sections were analyzed in the
hadronic mass range from 1.41 to 1.89 GeV. The overall $Q^{2}$-coverage ranges 
from 0.5 to 1.5 GeV$^2$. In the 2005 version of this analysis approach 
(JM05) \cite{Mo06-1,jm06},
double charged-pion 
production was described by the 
superposition of quasi-two-body channels with the formation and
subsequent decay of unstable particles in the intermediate states: 
$\pi^{-} \Delta^{++}$, $\pi^{+} \Delta^{0}$, $\rho^{0} p$, 
$\pi^{+} D^0_{13}(1520)$, $\pi^{+} F^0_{15}(1520)$, $\pi^{-} P^{++}_{33}(1640)$.
Remaining  direct 2$\pi$ production mechanisms without formation of unstable
intermediate particles were described by a set of exchange terms with the
amplitudes as outlined in \cite{Az05}.The production amplitudes for the first three quasi-two-body 
intermediate states were treated as sums
of $N^{*}$ excitations in the $s$-channel and non-resonant mechanisms
described in Refs. \cite{Mo01,Mo03,Mo07-1,jm06}. All well established 
resonances
with observed decays to the two pion
final states were included as well as $3/2^{+}(1720)$ candidate state that was 
observed in the analysis of 
2$\pi$ electroproduction data \cite{Ri03}.The production amplitudes for the 
$\pi^{+} D^0_{13}(1520)$, $\pi^{+} F^0_{15}(1685)$ 
and $\pi^{-} P^{++}_{33}(1640)$ intermediate state are
 described
in \cite{jm06}.

In the JM05 approach we succeeded in describing all
before mentioned observables in the CLAS data \cite{Ri03}. 
These results are presented 
in \cite{Mo06-1,jm06}.

\begin{figure}[htbp]
\vspace{17.0cm}
\includegraphics{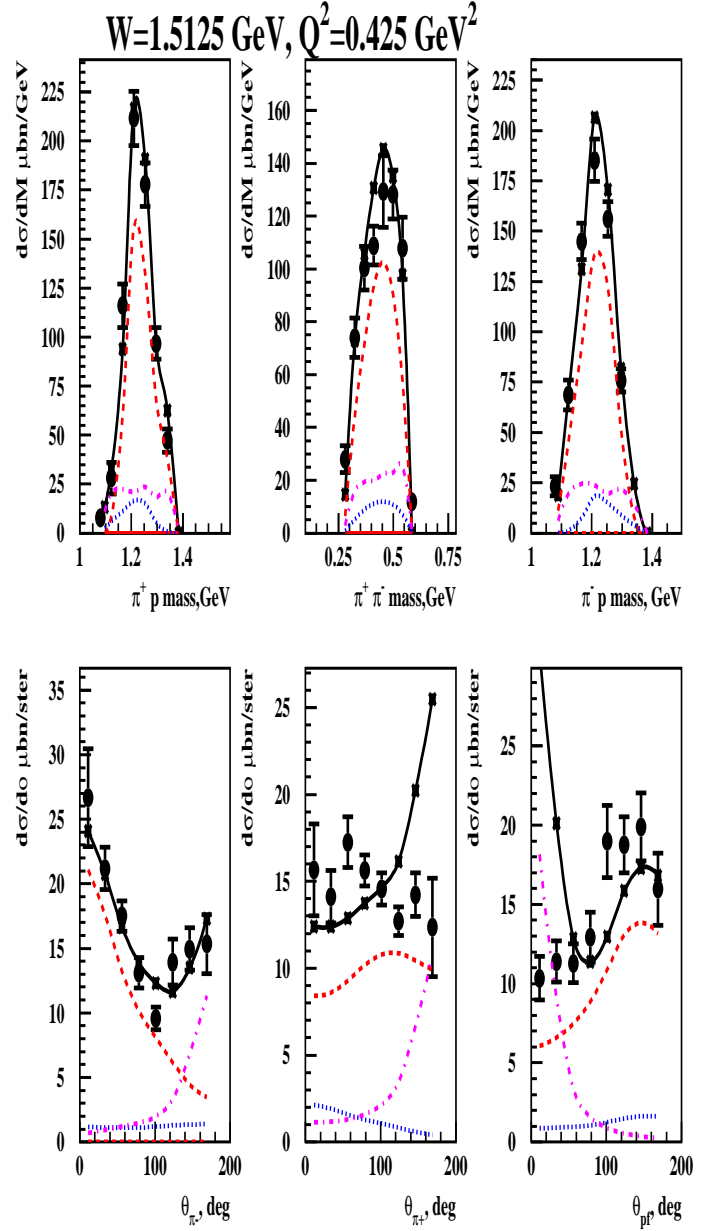} 
\caption[]{\small{Description of differential 2$\pi$ cross sections
at $W$=1.51 GeV and $Q^{2}$=0.425 GeV$^2$} in comparison within the framework
of JM05 (solid lines). The contributions from $\pi^{-} \Delta^{++}$, $\pi^{+}
\Delta^{0}$ isobar channels and direct 2$\pi$ production mechanisms are shown 
by dashed, dotted  and dot-dashed lines, respectively.  }
\label{failure}
\end{figure}

In the analysis of the most recent data at $W$$<$1.6 GeV and photon
virtualities from 0.2 to 0.6 GeV$^2$ \cite{Fe07,Mo07} for we
attempted the first time to
fit contributing mechanisms to the set of nine single-differential cross sections in each
($W$,$Q^{2}$) bin covered by measurements.  
In addition to the
differential cross sections mentioned above, they also included
 $\pi^+$ and $p$ 
angular distributions
and three distributions over angles $\alpha_{i,j}$ between two planes, composed 
by the momenta of
the initial proton and  final hadron (first plane) and two the other
final hadrons (a second plane) for three possible combinations 
amongst these pairs. We found that 
the JM05 model reasonably describes the data over all invariant masses and
$\pi^{-}$ angular distributions, as it is shown in Fig.~\ref{failure}. All these
observables were previously 
studied in
2$\pi$ electroproduction at higher photon virtualities \cite{jm06}. However,
JM05 model version failed in reproducing $\pi^{+}$ and $p$ CM
angular distributions included in the analysis for the first time. As it
follows 
from Fig.~\ref{failure}, where the contributions from
various mechanisms are presented, this failure is related to shortcomings in 
the description 
of direct 2$\pi$ production
mechanisms in \cite{Az05}.

\begin{figure}[htbp]
\vspace{5.0cm}
\includegraphics{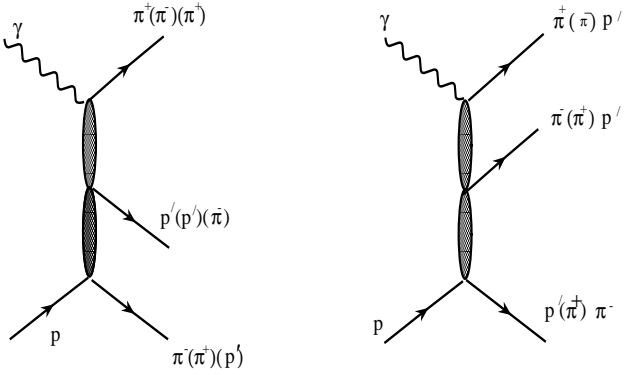} 
\caption[]{\small{Direct 2$\pi$ production mechanisms in JM06 model.}\label{diag}}
\end{figure}

In order to achieve a reasonable description of all angular distributions, we modified the
dynamics of direct 2$\pi$ production mechanisms with respect to those 
used in the JM05
version. 
The mechanisms of \cite{Az05} were substituted by ladder-type double 
exchange processes, shown in Fig.~\ref{diag}.

 The amplitudes of these processes are parametrized as Lorentz-invariant 
 contractions between
 spin-tensors of the initial and final particles, while the propagators for exchange
 mechanisms are described by exponents. We refer to this new approach as JM06
 model.
  In this model we succeeded
to describe recent CLAS data \cite{Fe07,Mo07} in the entire kinematics covered by
the measurements \cite{Fe07,Mo07}. As a typical example, the description of  
single-differential 2$\pi$
cross sections within the framework of the JM06 version is shown in 
Fig.~\ref{1diff} together with the contributions
from various mechanisms of JM06 approach.

The shapes of cross sections for various contributing mechanisms are
substantially
different in the observables, but highly correlated by mechanism
dynamics. Therefore, the successful description of all differential 
cross sections allowed us to pin down all major
contributing processes and access their dynamics at the phenomenological level.

To check the reliability of 
the amplitudes for contributing
processes, derived in this phenomenological data analysis, we fixed all
JM06 parameters, fitting them to  six single-differential cross sections$:$ all invariant
masses, and three final state angular distributions. The remaining three 
distributions 
over the $\alpha_{i,j}$ angles were calculated, keeping JM06 parameters fixed. 
A reasonable description of $\alpha_{i,j}$ angular
distributions was achieved in the entire kinematics covered 
by measurements. Therefore, we
confirmed the reliability of 2$\pi$ electroproduction mechanisms established 
in phenomenological data
analysis within the framework of the JM06 model.

\section{Cross sections for contributing mechanisms and $N^{*}$ electrocouplings.}
\label{sec:2} 
The contributions from isobar channels 
to the 2$\pi$ electroproduction, and the
electrocouplings for $P_{11}(1440)$ and $D_{13}(1520)$ resonances at $Q^{2}$ $<$ 0.6 GeV$^2$ 
were determined within the framework of the JM06 
approach. Electrocouplings of all
resonances were varied around their initial values, that were obtained by interpolating 
previous CLAS and world data. They were varied randomly  according to a 
normal distribution with a $\sigma$ of 30 \%. Simultaneously,
non-resonant mechanism parameters were varied with a $\sigma$ of 10 \%. 
For each trial set of
JM06 parameters we calculated nine differential cross sections in all ($W$,$Q^{2}$) bins,
covered in the CLAS measurements \cite{Fe07,Mo07}. Normalized to the amount of
data points $\chi^{2}/d.p.$
were
estimated from the comparison between measured and calculated differential 
cross sections. Finally, we selected calculated differential cross sections, 
which were closest to the experimental data applying the restriction 
for $\chi^{2}/d.p.$ listed in 
Table~\ref{chi2dp}. 

\begin{table}
\caption{$\chi^{2}$/d.p. achieved in the fit of recent CLAS 2$\pi$ data
\cite{Fe07,Mo07} within
the framework of JM06 model.}
\label{chi2dp}       
\begin{tabular}{lll}
\hline\noalign{\smallskip}
$Q^{2}$ interval, GeV$^2$ & $\chi^{2}$/d.p.    \\
\noalign{\smallskip}\hline\noalign{\smallskip}
0.25-0.40 & $<$ 2.8  \\
0.40-0.50 & $<$ 1.9  \\
0.50-0.60 & $<$ 1.8  \\
\noalign{\smallskip}\hline
\end{tabular}
\end{table}

\begin{figure}[htbp]
\vspace{14.0cm}
\includegraphics{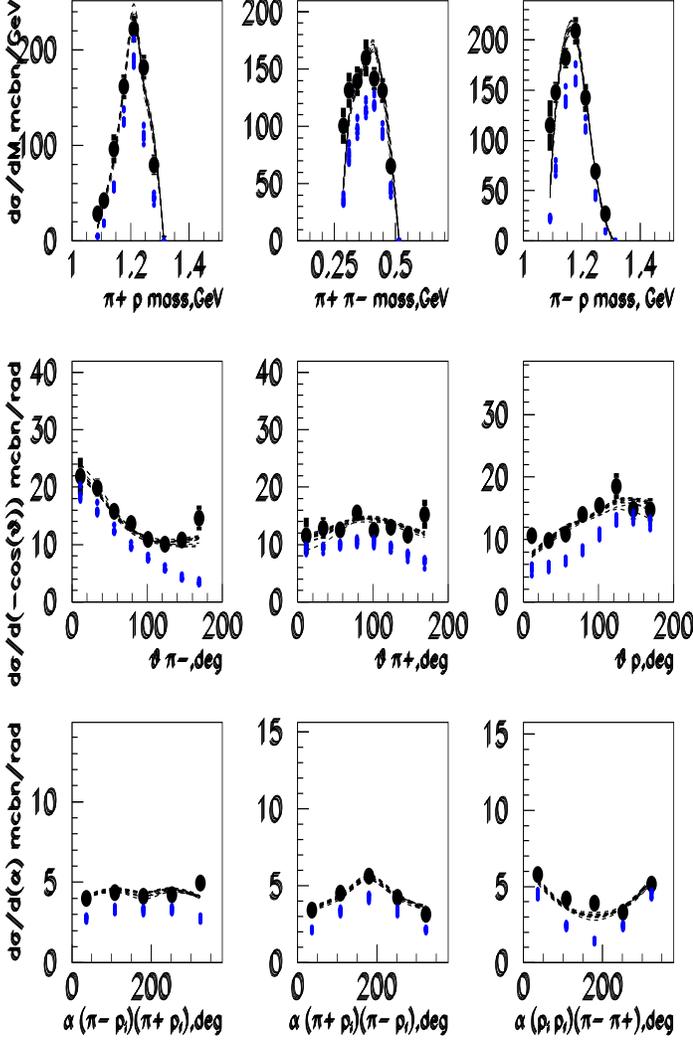} 
\caption[]{\small{Fit of differential 2$\pi$ cross sections within the framework of JM06 at
$W$=1.43 GeV and $Q^{2}$=0.425 GeV$^2$. Bunch of curves represent calculated cross section,
selected in fitting procedure (see Sect~\ref{sec:2})}. cross sections
corresponding to
contribution from all isobar channels combined are shown by vertical bars.}
\label{fitsec}
\end{figure}

The differential cross sections selected from the fit  at $W$=1.43 GeV and $Q^{2}$=0.425
GeV$^2$ are shown in Fig.~\ref{fitsec} by various dashed lines. For each 
calculated differential cross section,
selected in fitting procedure, we estimated the contribution from all 
isobar channels combined. The contribution from all isobar channels
combined is 
shown in  Fig.~\ref{fitsec} by vertical bars. The information on the isobar channel
contributions to 
all nine differential 2$\pi$ cross sections was obtained in our analysis for 
the first time. Fits within the framework of JM06 also enabled us to obtain 
amplitudes
for the superposition of all quasi-two-body processes mentioned in the 
Section~\ref{sec:1}, as well as for any individual isobar channel.

This information is of particular interest for
future $N^{*}$ studies  in a combined analysis of an
1$\pi$ and 2$\pi$ exclusive channels within the framework of an advanced coupled channel approach,
which is currently under development by EBAC \cite{Lee06}. Moreover, the data 
on isobar channel
differential cross sections and amplitudes open up new opportunities to 
establish explicit
meson-baryon mechanisms contributing to various isobar channels. Predictions
from various models, based on effective meson-baryon Lagrangians 
\cite{Lee06,Ki06,Na02} may be compared with 
isobar channel cross sections and amplitudes determined from the CLAS 2$\pi$ data analysis.

\begin{figure}[htbp]
\vspace{17.0cm} 
\includegraphics{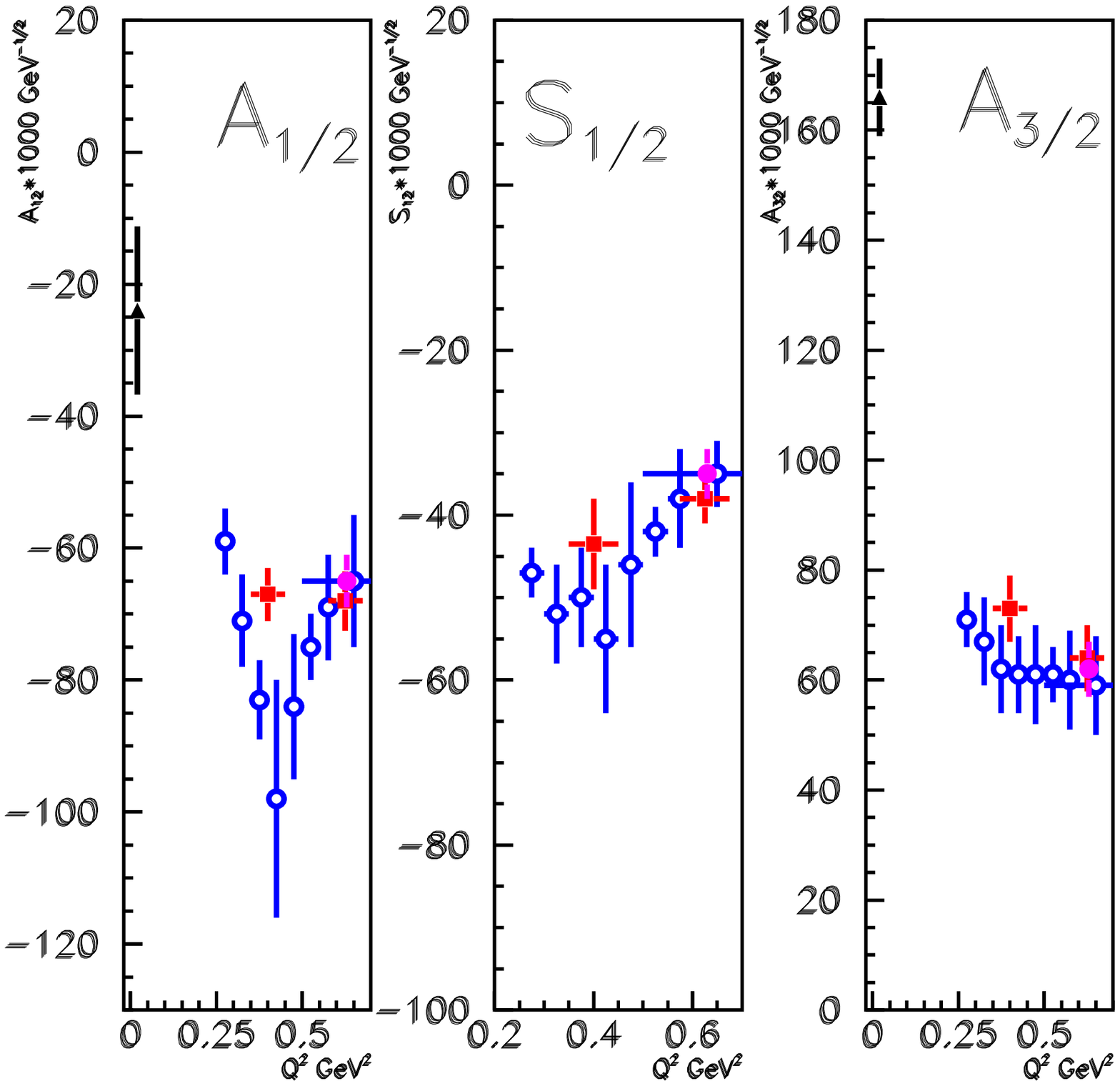}
\includegraphics{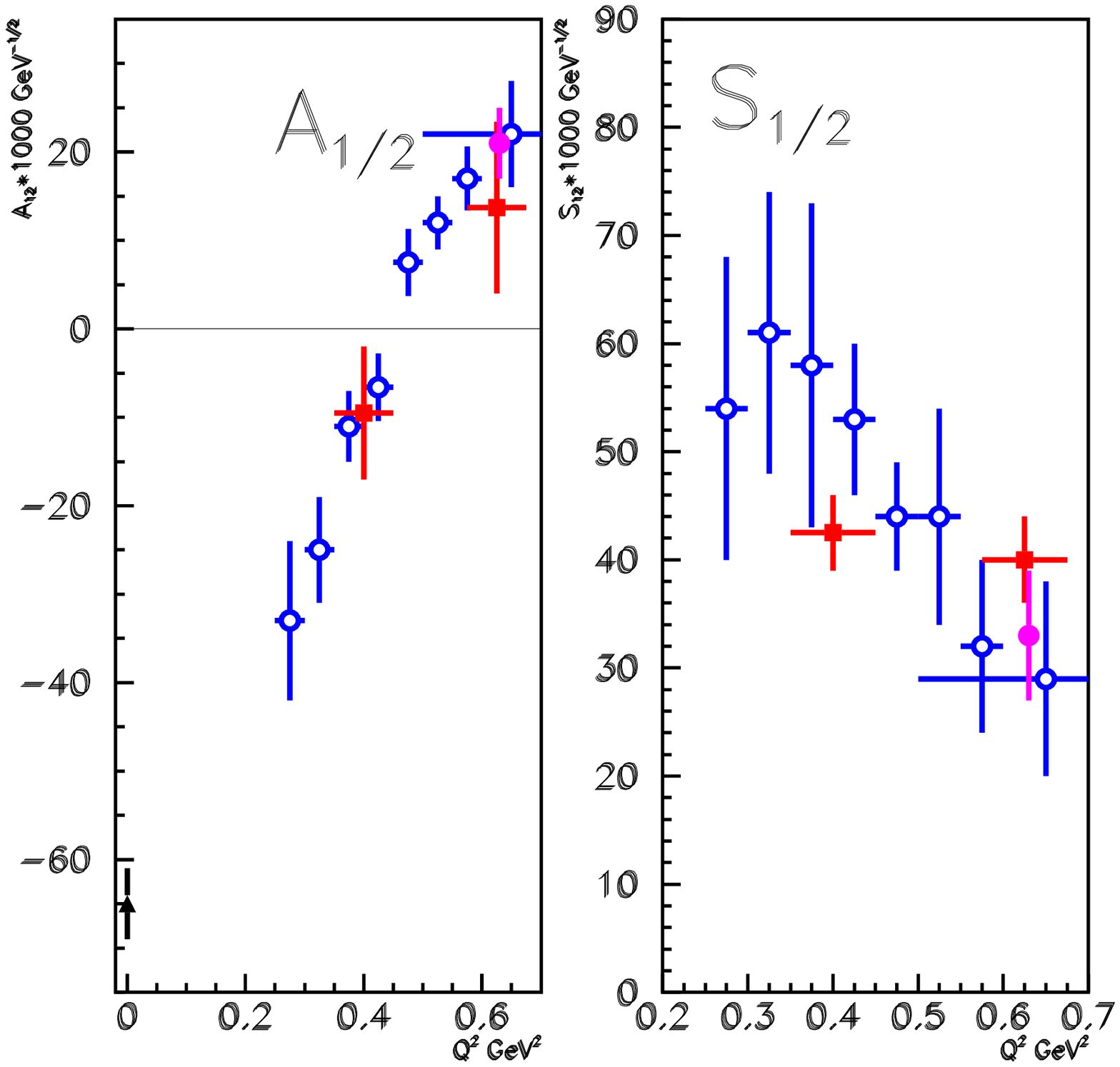} 
\caption[]{\small{Electromagnetic transition form factors for the $P_{11}(1440)$ (left) and  
$D_{13}(1520)$ (right) states, determined from analysis of CLAS single pion data 
(squares) and double pion data (open circles). The results from 
the combined 
1$\pi$/2$\pi$ data analysis at $Q^2$=0.65~GeV$^2$ are shown by filled
circles. The 
photocouplings from the PDG are shown by triangles.}}
\label{p11d13}
\end{figure}

Fig.~\ref{p11d13} shows the electrocouplings of the $P_{11}(1440)$ and 
$D_{13}(1520)$ states.  
For the 
first time, we obtain the $Q^{2}$ evolution of electrocouplings for these
states from the $\pi^{-}\pi^{+} p$ channel at
$Q^{2}$ from 0.2 to 0.6 GeV$^2$. 

These photon virtualities are particularly
sensitive to the contributions from $N^{*}$ meson-baryon dressing. The
electrocouplings
obtained from this analysis are in reasonable agreement
with the results from 1$\pi$ exclusive channel \cite{Az05b}, as well as from 
the combined
1$\pi$/2$\pi$ analysis \cite{Az05}. The consistency of the data on the 
$P_{11}(1440)$, $D_{13}(1520)$
electrocouplings, obtained from analysis of the two major 1$\pi$ and 2$\pi$ exclusive 
channels with substantially
different non-resonant processes demonstrates that a  reliable evaluation of 
these fundamental quantities can be obtained from the 1$\pi$ and 2$\pi$ 
electroproduction data. The analysis of the CLAS data on 
2$\pi$ electroproduction provides compelling evidence for the sign flip of the 
$A_{1/2}$
electrocoupling of the $P_{11}(1440)$ state at $Q^{2}$ in the range from 0.4 to
0.5 GeV$^2$ (Fig.~\ref{p11d13}).

\section{Conclusions and outlook.}

\begin{itemize}
\item The analysis of CLAS data on 2$\pi$
electroproduction \cite{Fe07,Mo07} allowed us to establish all significant mechanisms 
in this exclusive
channel at $W$ $<$ 1.6 GeV and $Q^{2}$ from 0.2 to 0.6 GeV$^2$. The JM06 model 
provides a 
reasonable description of all CLAS and world 2$\pi$ electroproduction data and 
may be 
used to separate resonant and non-resonant amplitudes in 
2$\pi$ electroproduction with the goal
of determining $N^{*}$ electrocouplings. 

\item The contributions to the double charged-pion electroproduction from 
isobar channels 
were established and will be used in future studies of
nucleon resonances in a
combined analysis of the CLAS data on 1$\pi$ and 2$\pi$ electroproduction within the framework
of an advanced coupled channel approach \cite{Lee06}.

\item The $P_{11}(1440)$ and $D_{13}(1520)$ electrocouplings were determined for the first time
from 2$\pi$ electroproduction data at $Q^{2}$ $<$ 0.6 GeV$^2$. This kinematic 
regime is
expected to be particularly sensitive to meson-baryon dressing effects in the nucleon
resonance structure. The analysis will be extended to obtain
electrocouplings for excited proton states with masses less then 2.5 GeV and 
in a large range of photon virtualities.

\item Comprehensive information on $N^{*}$ electrocouplings in a wide $Q^{2}$ 
range offers new
challenging opportunities for baryon structure theory to access fundamental mechanisms
responsible for baryon formation.
\end{itemize}

$\bf{Acknowledgements}$. The work was supported
by U.S. DOE contract DE-AC05-060R23177 under which
Jefferson Science Associates, LLC, operates the Jefferson Lab.

\end{document}